\documentclass[12pt]{article}
\usepackage{color}
\usepackage{epsf}
\usepackage{graphics}
\usepackage[dvips]{graphicx}
\usepackage{multirow}
\usepackage{amsmath}
\title{Isospin-symmetry in $pp$ and $AA$ collisions 
}

\author{A.V.~Guskov, G.I.~Lykasov, A.I.~Malakhov,
	A.A.~Zaitsev}
\begin{document}
	\maketitle
	\begin{center}
		{Joint Institute for Nuclear Research, 141980, Dubna,

			Moscow region, Russia}
	\end{center}
	\vspace{0.2cm}
	
	{\bf Abstract }
	
	The hypothesis of isospin-symmetry violation in pp and AA collisions
	is discussed. We show that the ratio
	of charged-to-neutral kaon cross sections, 
	$R_K~=~\frac{\sigma_{K^+}+\sigma_{K^-}}{2\sigma_{K^0_S}}$, 
	calculated within the similarity approach,
	including the gluon transverse momentum dependence (TMD) at low QCD scales, and using the 
	Lund model in the form of the MC generator PYTHIA, is above 1
	at energies $\sqrt{s}$ up to 
	20 GeV, even in the case of isospin-symmetry conservation.
	It decreases to 1, when the energy rises to a few TeV (ALICE data). 
	The reason for this is related to the dynamics of kaon production.
	This energy behavior of $R_K$ does practically not depend on the sort of 
	beam or target.

	\vspace{1.0cm}
	
	\section{Introduction}
	
	As it is well known, flavor symmetry exists in the quark sector, which means
	that interactions are independent of the quark type (flavor), when quarks
	are approximately massless. 
	For light quarks up ($u$) and down ($d$) this reduces to isospin symmetry. 
	The  difference $m_u-m_d$ is about 2 MeV, which is much smaller than the QCD
	scale $\Lambda_{QCD}\simeq$ 250 MeV, therefore, isospin symmetry is conserved for light quarks.
	
		The terms "isotopic spin" or "isospin" were suggested in
		\cite{Heisenberg:1932,Wigner:1937} explaining the similarity of the proton to
		the neutron, which can be due to their masses being approximately equal.
		This assumption has been extended to nuclei \cite{Werner;2006},
		analyzing the isospin-symmetry in collective nuclei.
		According to quantum mechanics, the Hamiltonian of the system,
		containing protons and neutrons is
		symmetric with respect to the exchange of these particles. In other words,
		the energy operator is invariant under this exchange.
		There exists a similar isospin-symmetry for systems containing 
		$u$ and $d$ quarks.
		This symmetry is the intrinsic
		property of the quark system structure, for example, the nucleon or the
		nucleus.
		However, the production of hadrons in $NN$ or $AA$ collisions is determined by
		the quark structure of the colliding particles and their 
		fragmentation into hadrons. 
		Therefore, for $K$-meson production in $NN$ or $AA$ collisions the ratio 
		$R_K=(\sigma_{K^+}+\sigma_{K^-})/
		(\sigma_{K^0}+\sigma_{\bar{K^0}})$ cannot be equal to 1.
		Assuming isospin symmetry for quarks the ratio $R_K$ can be equal to 1
		only when the fragmentation of quarks into hadrons is ignored.
		However, this fragmentation is a very important property of any dynamical
		approach analyzing multi hadron production in $NN,NA$ and $AA$ collisions
		at high energies. 
		It allows to obtain a realistic description
		of experimental data on hadron production in $NN$ or $AA$ collisions at high
		energies, see, for example \cite{Kaid1,Lund}.      
		
	
	Recently the assumption was made of the violation of isospin symmetry in
	$AA$ kaon production \cite{Isotp_vial}. This assumption is based on experimental
	data on the ratio $R_K~=~\frac{\sigma_{K^+}+\sigma_{K^-}}{2\sigma_{K^0_S}}$, which
	show that $R_K>$ 1 at 2-3 GeV$<\sqrt{s}<$200-300 GeV and $R_K=$ 1 at
	$\sqrt{s}=$ 3 TeV.
	
	In this paper we check the role of the dynamics of kaon production in $pp$ and
	$AA$ collisions, $pp\rightarrow K X$ and $AA\rightarrow K X$.
	
	\section{Kaon production in $pp$ and $AA$ collisions at low rapidity $y$ }
	
	We shall follow the similarity approach
	proposed in \cite{4,5} to hadron production in $AA$
	collisions at small transverse momenta.
	It is modified in \cite{LM:2018,ML:2020,LMZ:2021,LMZ:2024}. The conservation law of four-momenta
	is the following:
	
	\begin{equation}
		{(N_AP_A + N_BP_{B} - p_1)}^2 = 
		{(N_Am_0 + N_B m_0 + M)}^2 ,
		\label{eq:n2}
	\end{equation}
	where $N_A$ and $N_B$ are the respective fractions of the four-momentum
	transmitted by
	nucleus $A$ and nucleus $B$, the forms of $N_A, N_B$  are presented in
	\cite{4,5,LM:2018}; 
	$P_A$, $P_B$, $p_1$ are the four-momenta of nuclei  
	$A$,  $B$ and hadron $h$, respectively; $m_0$ is the mass of the nucleon; $M$ is
	the mass of the 
	particle providing conservation of the baryon 
	number, strangeness and other quantum numbers.
	It allows us to find the minimal value of $M$.   
	For $\pi$-mesons $m_1 = m_\pi$  and $M = $0.
	For antinuclei $M=m_1$ and for $K^-$-mesons  $M = m_1 = m_K$,
	$m_K$ is the mass of the $K$-meson.
	For nuclear fragments $M = - m_1$.
	For $K^+$-mesons $m_1 = m_K$ and $M = m_\Lambda  - m_0$,
	$m_\Lambda$ is the mass of the $\Lambda$-baryon.
	
	The inclusive spectrum of hadron $h$ produced in the $AB$ collision
	can be parameterized as a general universal function dependent on the 
	similarity parameter $\Pi$, as it was shown in \cite{Baldin_AA:1996}:
	\begin{equation}
		E d^3 \sigma_{AB}/d^3p~=~A_A^{\alpha(N_A)}\cdot A_B^{\alpha(N_{B})}\cdot F(\Pi)
		\label{eq:n4} 
	\end{equation}
	where $\alpha(N_A)=1/3 + N_A/3$, $\alpha(N_B)=1/3 + N_B/3$ 
	and function $F(\Pi)$ is the inclusive spectrum of hadron production in the $NN$
	collision . Here $F(\Pi)$ at $y=$ 0 has the same form as the inclusive spectrum
	$\rho_{NN}(y=0,p_t)$ \cite{BGLP:2012,GLLZ:2013}:
	\begin{equation}
		\rho_{NN}(y=0,p_T)~=~\rho_{q}(y=0,p_T)~+~\rho_{g}(y=0,p_T)~,
		\label{def:rhoNN}  
	\end{equation}
	with substitution of transverse momentum $p_T$ by $\Pi$
	\cite{LM:2018,LMZ:2024}.
	
	The function $\Pi$ has the following form \cite{4,5}
	\begin{equation} 
		\Pi=\min \frac{1}{2} \sqrt{(u_A N_A + u_B N_B)^2}  ,
		\label{eq:Pi} 
	\end{equation} 
	where $u_A$ and $u_B$ are the four-velocities of nuclei $A$ and $B$, $\Pi$ 
	is found from 
	the minimization of Eq.~\ref{eq:Pi} by solving the equation
	\cite{4,5} at $y=$ 0 and $N_A=N_B=N$:                                            
	\begin{eqnarray}
		\frac{d\Pi}{d N}=0
		\label{def:minimPi} 
	\end{eqnarray}  
	The exact solution of Eq.~\ref{def:minimPi} at $y=0$,
	as 
	\begin{eqnarray}
		N=\frac{\Pi}{\mbox{cosh}(Y)}\equiv\frac{2m_0\Pi}{\sqrt{s}}, 
		\label{def:N} 
	\end{eqnarray}  
	was obtained in \cite{4,5}, for details see, also \cite{LM:2018}. In Eq.~\ref{def:N} $Y$ is the rapidity of colliding nuclei.
	
	Therefore, $\alpha(N)=1/3 + 2m_0\Pi/(3\sqrt{s})$.
	Function $F(\Pi)$ has the following form \cite{LM:2018}:
	\begin{eqnarray}
		F(\Pi)=\bigg[ A_q \mbox{exp}\Big(-\frac{\Pi}{C_q}\Big) + \\
		\nonumber
		+
		A_g\sqrt{p_T}\phi_1(s) \mbox{exp}\Big(-\frac{\Pi}{C_g}\Big)\bigg] \sigma_{tot}
		\label{def:F} 
	\end{eqnarray}
	where 
	\begin{eqnarray}
		\Pi(s,m_{1T},y)~=~\left\{\frac{m_{1T}}{2m_0\delta_h}+
		\frac{M}{\sqrt{s}\delta_h}\right\}\mbox{cosh}(y)G,
		\label{eq:n10}
	\end{eqnarray}
	\begin{eqnarray}
		G~=~\left\{1+\sqrt{1+\frac{M^2-m_1^2}
			{(m_{1T}+2Mm_0/\sqrt{s})^2\mbox{cosh}^2(y)}\delta_h}\right\}~
		\label{def:G}.
	\end{eqnarray}
	Here 
	$\phi_1(s)~=~1-\sigma_{nd}(s)/\sigma_{tot}(s)$, see \cite{LM:2018,LMZ:2021},
	$\delta_h=\left(1 - \frac{s_{th}^h}{s} \right)$;
	$s_{th}^{\pi}\simeq 4m_0^2$;
	$s_{th}^{K^+}=\left(m_0 + m_K + m_\Lambda \right)^2$;
	$s_{th}^{K^-}=(2m_0+2m_K)^2$;
	$M = m_\Lambda - m_0; m_\Lambda = $ 1.115 GeV; 
	$m_K = $ 0.494 GeV; $s_0 = $ 1 GeV$^2$ is introduced to
	make $s/s_0$ dimensionless; $m_0 = 0.938$ GeV;
	$p_{1T}$ and $m_{1T}$ are the transverse momentum and transverse mass
	of the produced hadron $1$; 
	$\sigma_{nd} = (\sigma_{tot} - \sigma_{el} - \sigma_{SD})$ is the 
	non-diffractive cross-section;
	$\sigma_{tot},\sigma_{SD}$ and $\sigma_{el}$ are the total
	cross-section, the single diffractive cross-section and the elastic
	cross-section of $pp$ collisions, respectively.
	See details in \cite{LM:2018,LMZ:2021,LMZ:2024}. 
	\begin{figure*}[h!]
			\includegraphics[width=28pc]{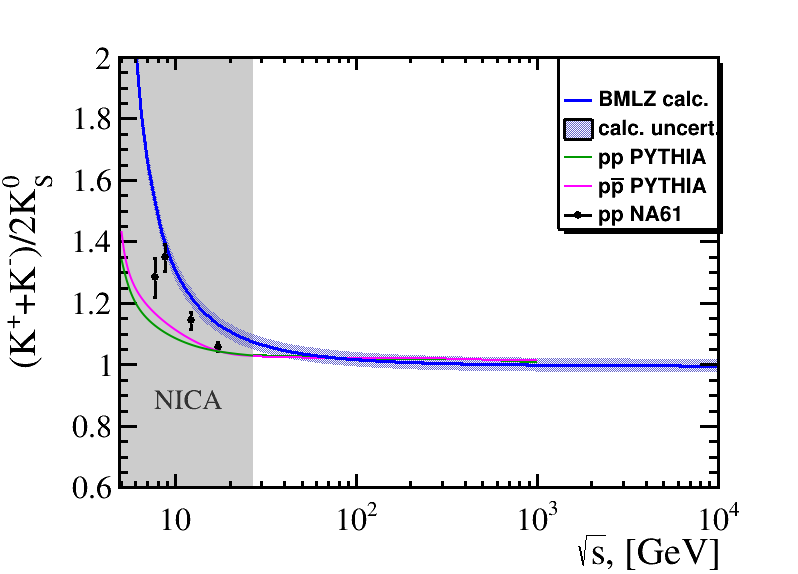}
			\caption{\label{fig1} The ratio
				$R_K~=~\frac{\sigma_{K^+}+\sigma_{K^-}}{2\sigma_{K^0_S}}$ for
				$pp$ and $p{\bar p}$ collisions, 
				as a function of $\sqrt{s}$ calculated within the similarity
				approach termed BMLZ \cite{LMZ:2024} (blue line) and the Lund
				model \cite{Lund}
				(green-$pp$, purple-$p{\bar p}$). The band in the blue line
				corresponds to the uncertainties of $R_K$ calculation for $pp$
				collisions. 
				The experimental data are taken from \cite{NA61/SHINE:2017,NA61-K0s,NA61-K0S-2}.
			}
		\end{figure*}
		\begin{figure*}[h!]
				\includegraphics[width=29pc]{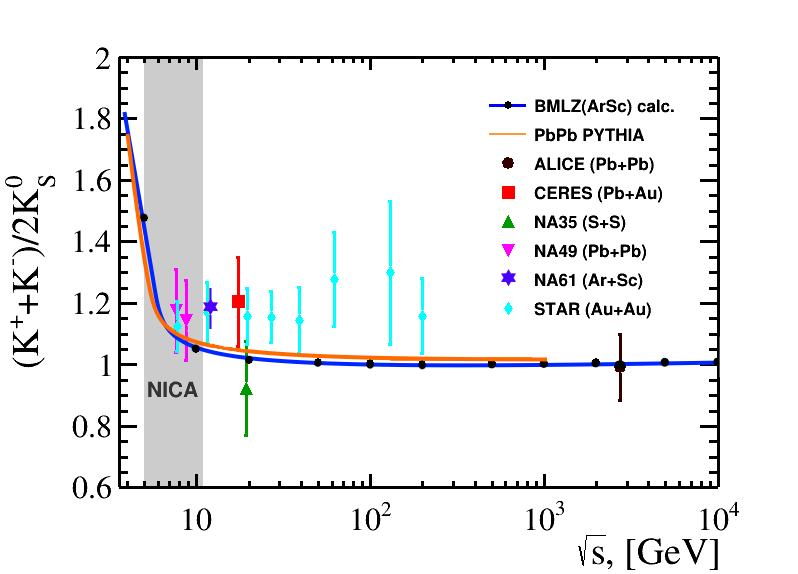}
			\caption{\label{fig2} The ratio
				$R_K~=~\frac{\sigma_{K^+}+\sigma_{K^-}}{2\sigma_{K^0_S}}$,
				as a function of $\sqrt{s}$ calculated within the similarity
				approach termed BMLZ for $ArSc$ collisions \cite{LMZ:2024} (blue line) and the Lund
				model \cite{Lund}
				(yellow line-for $PbPb$). 
				The experimental data are taken from \cite{Isotp_vial}.
			}
		\end{figure*}
		
		\begin{figure*}[h!]
				\includegraphics[width=30pc]{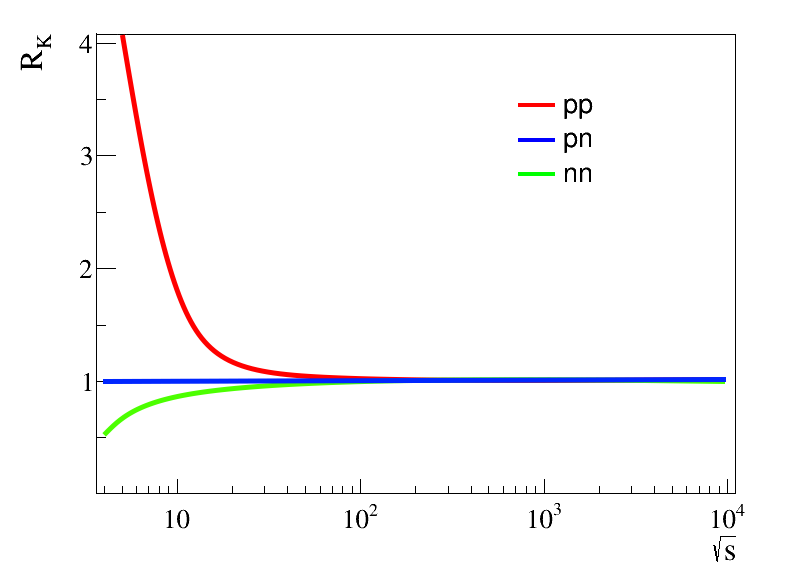}
		\caption{\label{fig3} The ratio
			$R_K~=~\frac{\sigma_{K^+}+\sigma_{K^-}}{2\sigma_{K^0_S}}$,
			as a function of $\sqrt{s}$ calculated within the similarity
			approach termed BMLZ \cite{LMZ:2024}. The red line corresponds to
			$R_K$ in $pp$, the blue line is $R_K$ in $pn$ and
			the green line is $R_K$ in $nn$ collisions.
		}
	\end{figure*}
	
	In Eq.~(7) the first term is the quark contribution and the second term
	is the gluon  contribution to the inclusive
	($NN\rightarrow h X$) spectrum \cite{BGLP:2012,GLLZ:2013}.
	Let us note that the gluon contribution to this inclusive spectrum is
	dominant as compared to the quark one at $p_T>$ 0.3 GeV$/$c.
	This dominance is due to the gluon transverse
	momentum distribution (TMD), see details in \cite{LLM:2023,LLM2:2023,LLM3:2023}.
	
	\section{Results and discussion}

	Calculating  the cross-sections of kaons in $pp$ and $ArSc$ collisions, according to Eqs.~(\ref{eq:n4},7),
	we compute the ratio $R_K~=~\frac{\sigma_{K^+}+\sigma_{K^-}}{2\sigma_{K^0_S}}$,
	as a function of $\sqrt{s}$. In Figs.~(\ref{fig1},\ref{fig2}) these ratios are presented for $pp$ and
	$ArSc$ collisions respectively within the BMLZ approach \cite{LMZ:2024} and for $pp,p{\bar p},pPb$
	collisions using the LUND model \cite{Lund}.
	These calculations are compared to the NA61/SHINE data Fig.~\ref{fig1},
	\cite{NA61/SHINE:2017,NA61-K0s,NA61-K0S-2}
	and the world data Fig.~\ref{fig2} at $x=$ 0 \cite{Isotp_vial}.
	One can see from Figs.~(\ref{fig1},\ref{fig2}) that the BMLZ and the Lund results of $R_K$ calculations give little bit different energy dependence
	for $pp$ and $AA$ collisions at $\sqrt{s}<$ 20 GeV, they match each other at $\sqrt{s}>$ 20 GeV and give $R_K=$ 1. It is due to the different
	$\sqrt{s}$ dependence of $R_K$ for $pp,pn$ and $nn$ collisions presented in Fig.\ref{fig3}. This difference reduces the ratio $R_K$ for
	$ArSc$ collisions compared to the $pp$ ones at $\sqrt{s}<$ 20 GeV. 
	
		Let us note that both approaches LUND and BMLZ take into account the
		fragmentation of all the quarks into hadrons. These fragmentation effects
		are sizable at initial energies close to the threshold
		of kaon 
		production in the case of different channels, involving minimal
		energy, like $K\Lambda N$ or $K^+ K^-NN$, for which $M=m_\Lambda - m_n$ and $M=m_K$.
		These minimal channels for different processes $NN\rightarrow K X$ are presented
		in the Appendix, Table 1.  
		
		The large contributions of these channels at low $\sqrt{s}$
		leads to an increase of the ratio $R_K $ above 1 as the
		initial energy $\sqrt{s}$ decreases, which is seen in Figs.~(\ref{fig1},\ref{fig2}).
		When $\sqrt{s}$ increases,the contributions of these hadron
		channels decrease, because of many other channels
		existing. Therefore,
		asymptotically at large $\sqrt{s}$ these initial channels do not contribute
		sizeably to the ratio $R_k$, which becomes close to 1, see
		Figs.~(\ref{fig1},\ref{fig2}) at
		$\sqrt{s}>$ 100 GeV. 
	The calculations of $R_K$ within these models have only taken into account
	the nuclear PDF ignoring the another nuclear effects, rescatterings {etc}.
	
	\section{Summary }
	
	Using the inclusive $p_T$-spectra of kaons and their production cross sections
	in $pp$ and $ArSc$ collisions calculated in \cite{LMZ:2021} and
	\cite{LMZ:2024}\,
	respectively, we have computed the ratio
	$R_K~=~\frac{\sigma_{K^+}+\sigma_{K^-}}{2\sigma_{K^0_S}}$, as a function
	of $\sqrt{s}$.
	The similar results for $R_K$ were obtained within the LUND model \cite{Lund} in the form of the MC generator PYTHIA.  
		Our calculations  of $R_K$ for $p~p$ collisions at $\sqrt{s}=$ 6-20 GeV do not contradict the NA61/SHINE data 
		\cite{NA61/SHINE:2017,NA61-K0s,NA61-K0S-2}
		and they have shown the decrease of $R_K$ from 2.0 (BMLZ) or 1.35-1.4 (PYTHIA) up to 1. Therefore, for $pp$ collisions
                the isospin-symmetry may not be violated. 
		
		Predictions of this ratio for $AA$ collisions at $\sqrt{s}=$ 6-10 GeV have shown  the decrease of $R_K$ from 1.35 up to 1.05, however,
		the NA49 (Pb~Pb) and STAR (Au~Au) data with big error bars are decreasing very slowly from 1.2 up to 1.15 at this energy region.
		At $\sqrt{s}>$ 10 GeV the NA61/SHINE, CERES and STAR data have illustrated that $R_K\simeq $1.2-1.3 with big error bars, about 20\%-30\% .
		Only NA35 (S~S) experiment has confirmed that $R_K\simeq$ 0.8-0.9 with error bars about 20\%-25\%.
		These predictions for $R_K$ at $\sqrt{s}=$ 6-20 GeV in $pp$ collisions and in $AA$ collisions at $\sqrt{s}=$ 6-10 GeV can be checked in
		future measurements at the NICA complex.

		The difference between our calculations and the world data at 11 GeV $<\sqrt{s}<$ 200 GeV can be related to our approximation,
		which does not take into account the nuclear effects, or to the isospin symmetry violation in $A~A$ collisions. Despite to this it is desirable to check
		the world data on $R_K$ in $A~A$ collisions by improving the measurement accuracy.

	\section{Appendix}
		The Table 1 illustrates the minimal energy channels for $NN\rightarrow K X$
		in column 4 and the corresponding $M$ values in column 5
		(Eqs.~\ref{eq:n10} and \ref{def:G}).
	
	\begin{table}[h!]
		\centering
		\caption{The minimal energy channels for K-meson
			production}
		\renewcommand{\arraystretch}{1.2}
		\begin{tabular}{|c|c|c|c|c|}
			\hline
			Type & № & Reaction & Min.\ channels & Mass \\
			\hline
			\multirow{4}{*}{$pp$}
			& 1 & $pp \to K^{+} + X$ & $K^{+}\Lambda p$ & $m_{\Lambda} - m_{N}$ \\
			\cline{2-5}
			& 2 & $pp \to K^{-} + X$ & $K^{+}K^{-}pp$ & $m_{K}$ \\
			\cline{2-5}
			& \multirow{2}{*}{3} & \multirow{2}{*}{$pp \to K^{0} + X$} & $K^{0}\,\bar{K}^{0}\,pp$ & $m_{K}$ \\
			\cline{4-5}
			&   &   & $K^{0}\Lambda p\,\pi^{+}$ & $m_{\Lambda} - m_{N}$ \\
			\hline
			\multirow{3}{*}{$pn$}
			& 4 & $pn \to K^{+} + X$ & $K^{+}\Lambda n$ & $m_{\Lambda} - m_{N}$ \\
			\cline{2-5}
			& 5 & $pn \to K^{-} + X$ & $K^{-}K^{+}pn$ & $m_{K}$ \\
			\cline{2-5}
			& 6 & $pn \to K^{0} + X$ & $K^{0}\overline{K^{0}}pn$ & $m_{K}$ \\
			\hline
			\multirow{4}{*}{$nn$}
			& \multirow{2}{*}{7} & \multirow{2}{*}{$nn \to K^{+} + X$} & $K^{+}K^{-}nn$ & $m_{K}$ \\
			\cline{4-5}
			&   &   & $K^{+}\Lambda n\pi^{-}$ & $m_{\Lambda} - m_{N}$ \\
			\cline{2-5}
			& 8 & $nn \to K^{-} + X$ & $K^{+}K^{-} nn$ & $m_{K}$\\
			\cline{2-5}
			& 9 & $nn \to K^{0} + X$ & $K^{0}\overline{K^{0}}nn$ & $m_{K}$ \\
			\hline
		\end{tabular}
	\end{table}

	{\bf Acknowledgments.}
	
	\begin{sloppypar} 
		We are very grateful to H.~Jung, M.~Gazdzicki, F~.Giacosa, O.V.~Teryaev for extremely
		helpful discussions.   
	\end{sloppypar}

\end{document}